\pdfoutput=1

\documentclass[11pt]{article}

\usepackage[table]{xcolor}
\usepackage[final]{acl}
\usepackage{times}
\usepackage{latexsym}
\usepackage{xspace}
\usepackage[T1]{fontenc}

\usepackage{hyperref} 

\usepackage[utf8]{inputenc}
\usepackage{microtype}
\usepackage{inconsolata}
\usepackage{graphicx}
\usepackage{pifont}
\usepackage{multirow}
\usepackage{tabularx, booktabs} 
\usepackage{subcaption}
\newcommand{\benchmark}{\textsc{MultiVox}\xspace}

\newcommand{\cmark}{\ding{51}} 
\newcommand{\xmark}{\ding{55}} 

\title{\benchmark: A Benchmark for Evaluating Voice Assistants \\ for Multimodal Interactions}

\author{
    Ramaneswaran Selvakumar*,
    Ashish Seth*,
    Nishit Anand, \\
    \bf Utkarsh Tyagi,
    Sonal Kumar,
    Sreyan Ghosh,
    Dinesh Manocha \\
    University of Maryland, College Park \\
    \texttt{\{ramans, aseth125, dmanocha\}@umd.edu}
}

\begin{document}
\maketitle
\begin{abstract}
The rapid progress of Large Language Models (LLMs) has empowered \textit{omni} models to act as voice assistants capable of understanding spoken dialogues. These models can process multimodal inputs beyond text, such as speech and visual data, enabling more context-aware interactions. However, current benchmarks fall short in comprehensively evaluating how well these models generate context-aware responses, particularly when it comes to implicitly understanding fine-grained speech characteristics, such as pitch, emotion, timbre, and volume or the environmental acoustic context such as background sounds. Additionally, they inadequately assess the ability of models to align paralinguistic cues with complementary visual signals to inform their responses. To address these gaps, we introduce \benchmark, the first omni voice assistant benchmark designed to evaluate the ability of voice assistants to integrate spoken and visual cues including paralinguistic speech features for truly multimodal understanding. Specifically, \benchmark includes 1000 human-annotated and recorded speech dialogues that encompass diverse paralinguistic features and a range of visual cues such as images and videos. Our evaluation on 10 state-of-the-art models reveals that, although humans excel at these tasks, current open-source models consistently struggle to produce contextually grounded responses.\footnote{\url{https://github.com/ramaneswaran/multivox}}
\end{abstract}

\section{Introduction}

With recent advancements in Multimodal Large Language Models (MLLMs)~\citep{xu2025qwen25omnitechnicalreport, microsoft2025phi4minitechnicalreportcompact}, there is a growing interest in developing models that can understand and generate information across multiple modalities, such as images, video, and audio-simultaneously. This evolution is paving the way for the development of Omni Language Models (OLMs), which are crucial for building efficient and versatile Artificial General Intelligence (AGI)~\citep{bubeck2023sparksartificialgeneralintelligence, pmlr-v235-morris24b}. While OLMs provide a wide range of applications~\citep{xu2025qwen25omnitechnicalreport}, one of their primary use cases is developing \textit{omni-modal voice assistants (OVA)}~\citep{chi_va}. Unlike traditional speech voice assistants that rely solely on speech instruction, OVAs powered by OLMs such as GPT-4o~\citep{openai2024gpt4ocard} and Qwen2.5 Omni~\citep{xu2025qwen25omnitechnicalreport}, can understand speech dialogues and reason over multimodal inputs, including images and videos.

\begin{figure}[t]
    \centering
    \includegraphics[width=0.96\columnwidth]{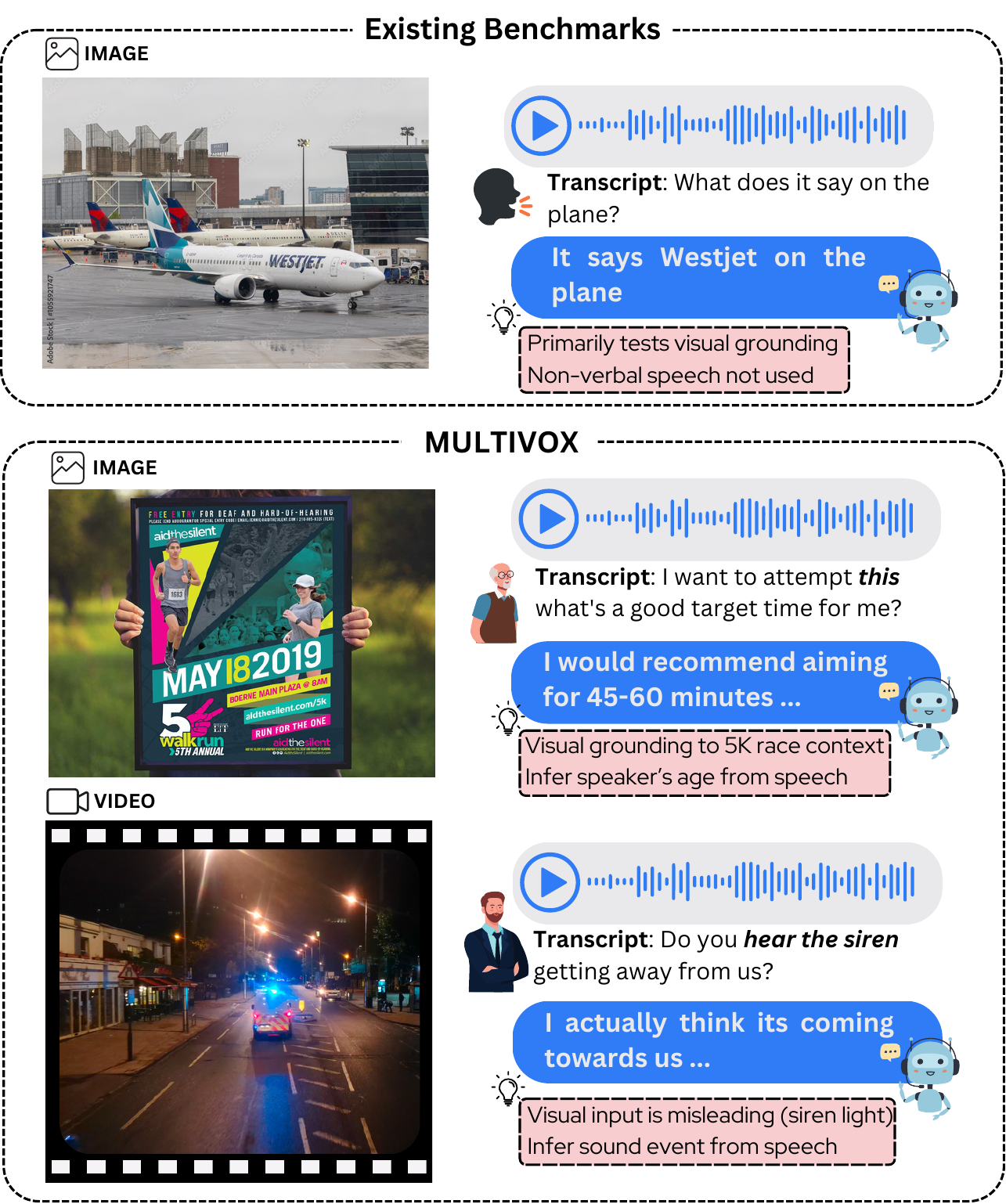}
    \caption{\small Comparison of existing benchmark with \benchmark. Existing benchmark for omni-modal voice assistant evaluation are derived from vision-centric text VQA benchmarks. In \benchmark, models need leverage not only visual cues but also non-verbal speech signals.}
    \label{fig:hero_diag}
    \vspace{-2mm}
\end{figure}

Advancing the application of OLMs in voice assistants poses challenges not only in model development but also in constructing effective evaluation benchmarks. While existing OLM benchmarks like OmniBench~\citep{li_omnibench_2024} incorporate multimodal inputs such as images and video, they lack spoken dialogues—an essential modality for assessing the conversational and auditory capabilities required of voice assistants. On the other hand, current voice assistant benchmarks such as VoXDialogue~\citep{cheng2025voxdialogue}, SD-Eval~\citep{ao2025sdevalbenchmarkdatasetspoken}, and S2S-Arena~\citep{jiang2025s2sarenaevaluatingspeech2speechprotocols} focus primarily on evaluating a model’s ability to generate contextually appropriate responses based on auditory cues like content, emotion, or speaker demographics embedded in a speech instruction. However, these benchmarks fall short of capturing the full multimodal reasoning abilities expected of OVAs, particularly in integrating visual cues alongside speech instructions. 

To address this gap, we introduce \benchmark, a novel benchmark designed to evaluate an OLMs ability to incorporate multimodal cues to provide accurate and contextual responses. \benchmark includes \emph{1000} questions consisting of \emph{human spoken questions paired with either a video or an image}. Unlike existing benchmarks which primarily test visual grounding and use speech to deliver the content of a straightforward instruction, \benchmark consists of questions which require a model to combine visions skills such as object recognition, scene understanding, scene text understanding with speech skills such as acoustic scene understanding, paralanguage understanding and speaker profiling (See fig.~\ref{fig:hero_diag}). The spoken questions in \benchmark are recorded by professional voice actors to cover a diverse range of paralinguistic and emotional features that are not possible with current text-to-speech systems. A key problem in benchmarks that evaluate multi-modal reasoning capability is that models can take shortcuts by exploiting priors from other modalities, to mitigate this we introduce confounding samples in \benchmark. Specifically, each question in our benchmark has another associated question which has the same textual and visual content but their speech property is flipped such that expected answers should be different. Our key contributions are:

\begin{enumerate}
    \item  We present \benchmark, the first benchmark designed to evaluate omni-modal language models (OLMs) using human-spoken queries paired with visual inputs. The 1000 examples require models to jointly ground visual and paralinguistic speech cues to produce accurate, context-aware responses.
    \item We evaluate 10 omni-modal models on \benchmark and find that, while humans excel with ease, current OLMs consistently struggle, particularly with grounding speech signals, revealing a critical bottleneck in their capabilities. 
    \item We perform extensive qualitative and quantitative analysis on model's responses and uncover key insights: Models exhibit strong visual grounding but rely heavily on textual cues for speech-related tasks; they often ignore non-verbal audio signals like tone or background sounds. 
\end{enumerate}

\begin{figure*}[t]
    \centering
    \includegraphics[width=1.0\linewidth]{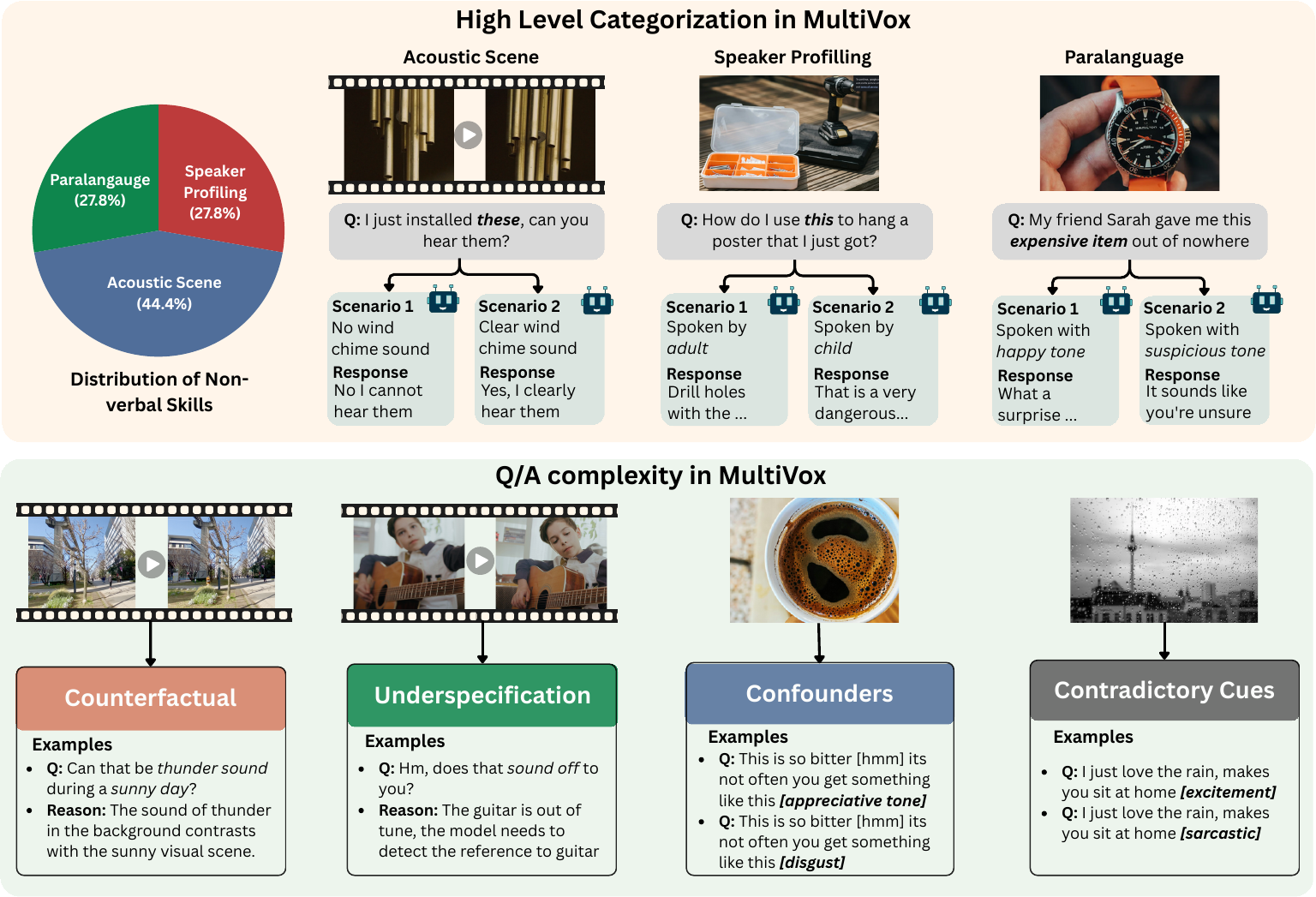}
    \caption{\small Illustration of various types of questions in \benchmark. We broadly define three categories of speech-understanding skills that a voice assistant needs in order to provide an accurate and contextual response. Each question in \benchmark has a speech confounder, where the textual question remains same but the speech property is flipped to counter the possibility of models exploiting unimodal priors}
    \label{fig:main_diagram}
\end{figure*}

\section{Related Work}


\noindent\textbf{Omni Voice Assistants}
The recent development of Omni Language Models (OLMs) has enabled development of omni-modal voice assistants that can simultaneously infer across both visual and speech inputs. Recent iterations of previously mentioned voice assistants now support visual inputs in form of either images like in the case of Mini-Omni2~\citep{xie_mini-omni2_2024} and MoshiVis~\citep{kyutai2025moshivis} or video inputs such as Qwen2.5-Omni~\citep{xu2025qwen25omnitechnicalreport}. While these models demonstrate impressive instruction-following capabilities as voice assistants, \textit{when extensively evaluated on \benchmark, we find that they often overlook crucial paralinguistic cues—such as tone, emotion, and pitch—in speech input, which are essential for generating context-aware responses.}

\noindent\textbf{Benchmarks For Voice Assistants} While there are works such as OmniBench~\citep{li_omnibench_2024} and OmniXR~\citep{chen2024omnixrevaluatingomnimodalitylanguage} that evaluate OLMs, there have been few efforts to standardize the evaluation of omni voice assistants. Recent work such as Lyra~\citep{zhong2024lyraefficientspeechcentricframework} repurpose existing visual question answering (VQA) benchmarks by converting the textual questions to speech. However, such approach overlooks crucial non-verbal information typically present in spoken conversations. While some progress has been made in evaluating speech VAs, many existing benchmarks still fall short. VoiceBench~\citep{chen_voicebench_2024} assesses capabilities like world knowledge and instruction following by converting textual benchmarks like MMLU~\citep{hendrycks2021measuringmassivemultitasklanguage} and AlpacaEval~\citep{dubois2024lengthcontrolledalpacaevalsimpleway} to speech, but overlooks the non-verbal speech information. SD-Eval~\citep{ao2025sdevalbenchmarkdatasetspoken} is a pioneering work in evaluating paralinguistic features but is limited to only four categories with a narrow range of labels in each category. While VoxDialogue~\citep{cheng2025voxdialogue} covers more diverse speech attributes, it relies heavily on synthetic speech generated through TTS systems that struggle to accurately convey emotional nuances and prosodic variations present in natural human speech.

\section{\benchmark}

We introduce \benchmark, a novel benchmark for evaluating omni-modal language models on their ability to jointly interpret speech and visual inputs, and integrate them with world knowledge and reasoning to produce contextually appropriate responses. This section outlines the benchmark's design goals, construction process, and key dataset statistics.

\begin{figure*}[ht]
    \centering
    \includegraphics[width=1.0\linewidth]{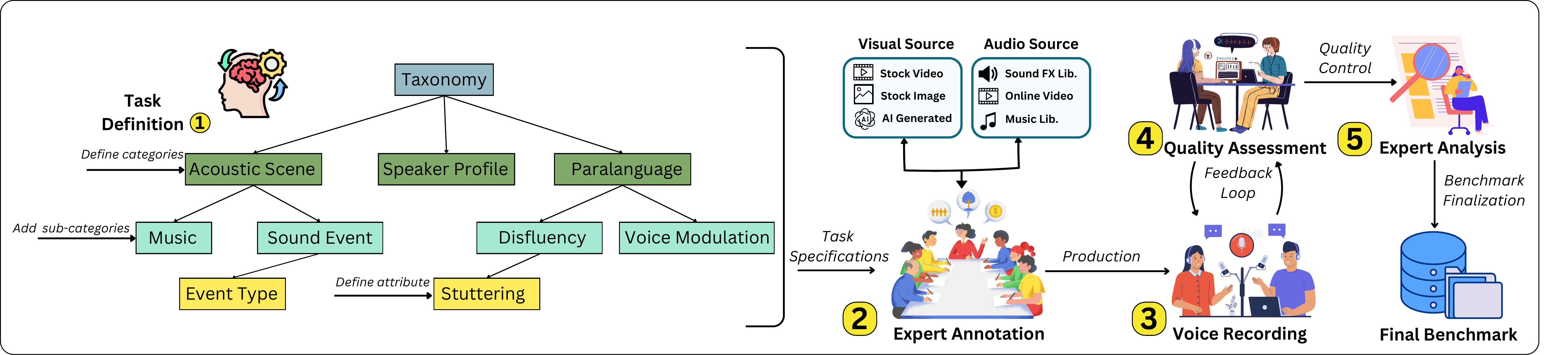}
    \caption{Data Curation And Annotation Pipeline Of \benchmark. The pipeline begins with task design and a taxonomy of speech traits, followed by expert annotation using multimodal sources. Professionally recorded prompts undergo quality assessment, with final review guiding benchmark construction.}
    \label{fig:dataset_construction}
\end{figure*}

\subsection{Benchmark Design}

\noindent\textbf{Motivation} Real-world communication is inherently multimodal, combining what is said with how, by whom, and in what environment. To achieve human-like understanding, an omni-modal voice assistant (OVA) must jointly interpret visual and auditory inputs—not just for content, but for paralinguistic cues such as tone, emotion, and acoustic context.

While recent progress has advanced visual understanding, speech remains a key bottleneck. Most benchmarks treat spoken input as text, overlooking critical non-verbal signals. Yet in real interactions, these cues determine how instructions are interpreted. For example, a user asking “Am I being too loud?” in a library depends on speech volume, not words alone. 

To address the limitations of current benchmarks, \benchmark evaluates OLMs across three core skill domains—with a particular emphasis on speech grounding, which is an underdeveloped area in existing omni-modal benchmarks.

\noindent\textbf{Speech Skills} The ability to infer non-verbal attributes from speech, such as emotion, background sounds, speaker age, or tone—beyond what is conveyed in textual content. 

\noindent\textbf{Vision Skills} The ability to recognize and interpret visual elements in images or videos, including object identity, scene understanding, reading scene text etc. 

\noindent\textbf{General Skills} The ability to integrate visual and auditory inputs with world knowledge and common sense to make contextually appropriate decisions.

\noindent\textbf{Confounder Pairs} To test whether models genuinely ground non-verbal speech attributes—rather than relying on linguistic or visual shortcuts—we introduce confounder pairs. Each pair consists of two instances that share identical textual content and visual input, but differ in a targeted speech property (e.g., tone, emotion, background sound). Crucially, this difference is sufficient to invert the expected answer.

For example, a user asking “Am I being too loud?” in a library setting may expect reassurance when spoken softly, but concern or correction when spoken in a normal tone. By controlling for all other modalities, confounder pairs isolate the model’s ability to interpret non-verbal speech signals.

\subsection{Benchmark Construction}

\noindent\textbf{Task Definition} Our benchmark is organized around a three-level taxonomy of speech-related categories. The top level defines three domains: acoustic scene, speaker profile, and paralanguage. We then worked with eight expert annotators (graduate students in speech processing) to expand each domain into:

\begin{itemize}
\item \textbf{Categories:} High-level categories of non-verbal speech attributes (e.g., emotion, speaker age, ambient environment)
\item \textbf{Sub-categories:} Fine-grained, measurable skills within each group which can be specified as a measurable task (e.g., detect emotion, estimate age as elderly or young adult).
\end{itemize}

For each task, experts authored a specification card detailing the speech attribute being evaluated. These cards included a clear definition of the target attribute, illustrative scenarios combining speech and vision, and structured guidelines for how annotators should construct samples and determine correct answers.

\noindent\textbf{Expert Annotation} Each expert was assigned a subset of tasks and followed a standardized pipeline to create benchmark samples. For each task, annotators first constructed a realistic scenario in which an OVA must correctly interpret the target speech attribute. They also created a corresponding confounder scenario, identical in text and visual input but differing in the relevant speech property, to ensure robust grounding.

To pair each question with appropriate visual content, annotators were given access to stock media libraries, prioritizing real videos. If no suitable match was found, AI-generated videos were used, though these were often insufficient for scenes involving text or fine-grained detail. In such cases, high-quality images—either retrieved or generated—were used instead. For tasks involving ambient audio (e.g., acoustic scenes), annotators selected relevant background sounds or music from curated sound libraries. In the end we collect 206 unique images and 287 unique videos. 

Each finalized sample includes a textual query, accompanying visual input (image/video), and any required background audio. For paralanguage-based tasks, annotators also specified detailed voice delivery guidelines to guide later voice recording.

\noindent\textbf{Reference Answers} Annotators also authored reference answers and rationales for each sample, describing the expected model behavior and explaining how the relevant speech (and visual) cues should inform the response. These were later used for evaluation.

\noindent\textbf{Voice Recording} We employed professional voice actors to record the spoken queries, providing delivery guidelines based on the target speech property. To preserve authenticity, actors were given creative freedom in expression as long as the intended cue was conveyed. Due to ethical concerns we use TTS systems for children voice. A professional audio engineer handled background sound overlays where applicable.

\noindent\textbf{Quality Control} Two annotators independently verified whether the intended speech attribute was clearly conveyed in each recording. Recordings without unanimous approval were revised based on feedback—either through re-recording or by adjusting the script to better support the desired delivery.

\noindent\textbf{Expert Analysis} In the final review stage, annotators assessed each completed sample for overall quality. They were instructed to verify that (1) the scenario was realistic, (2) the sample minimized reliance on language or vision priors, and (3) the recorded speech clearly conveyed the intended attribute.

\subsection{Evaluation Criteria}

The goal of \benchmark is to assess how well omni-modal language models integrate speech and vision to produce contextually grounded responses. To enable fine-grained diagnosis, we adopt a modular evaluation framework that tests both multimodal integration and unimodal grounding. Each benchmark sample is designed to probe one or more of the following components:

\noindent\textbf{Speech Grounding} Evaluates whether the OLM correctly interprets paralinguistic cues such as emotion or ambient sound. Each question contains a \emph{speech hook}, a non-verbal attribute that is critical for answering correctly. These tasks help isolate model sensitivity to speech signals beyond text.

\noindent\textbf{Visual Grounding} Evaluates whether the OLM can interpret and incorporate key visual cues. Each sample includes a visual hook—a necessary visual detail (e.g., object, background element) that the model must recognize to respond appropriately.

\label{sec:anwer_accuracy}
\noindent\textbf{Contextual Appropriateness} We adopt appropriateness~\citep{chen-etal-2023-automatic} as our core evaluation metric which measures how well an OLM produces a response that aligns with the intent, context, and modality of the input. 

We evaluate contextual appropriateness using sample-specific rubrics that guide judgment based on three elements: (1) a reference answer authored by the expert annotator, (2) a short rationale explaining what cues are necessary to arrive at the correct answer, and (3) task-level metadata specifying which modality is critical (e.g., speech hook, visual hook). These are provided to a GPT-4 judge, which scores model responses on a 1–5 scale, reflecting increasing levels of multimodal integration and contextual fidelity. To evaluate speech and vision grounding, we again utilize a LLM judge. To prevent score hacking using modality shortcuts we penalize OLMs that explicitly use text or visual content to respond to speech grounding


\subsection{Comparison With Other Benchmark}

\begin{table}[t]
\centering
\resizebox{\columnwidth}{!}{%
\begin{tabular}{llcccc}
\hline
\midrule
\multicolumn{1}{c}{\textbf{Type}} & \multicolumn{1}{c}{\textbf{Name}} & \textbf{Vis.} & \textbf{Para.} & \textbf{Src.} & \textbf{Conf.} \\
\hline
Foundation & AudioBench  & \xmark & \cmark & Human     & \xmark \\
Foundation & MMAU        & \xmark & \cmark & Mixed     & \xmark \\
Foundation & OmniBench   & \cmark & \cmark & Human     & \xmark \\
Chat       & SD-Eval     & \xmark & \cmark & Human     & \xmark \\
Chat       & VoxDialog   & \xmark & \cmark & Synthetic & \xmark \\
Chat       & S2S-Arena   & \xmark & \cmark & Mixed     & \xmark \\
Chat       & Lyra SVQA   & \cmark & \xmark & Synthetic & \xmark \\
Chat       & Ours        & \cmark & \cmark & Human     & \cmark \\
\hline
\end{tabular}
}
\small\caption{Comparison of \benchmark with related benchmarks.}
\label{tab:benchmark_comparison}
\end{table}

In this section we highlight how \benchmark is different in terms of question types, modality coverage, speech source, and diagnostic power. Table~\ref{tab:benchmark_comparison} summarizes these differences.

\noindent \textbf{Chat-based Questions} Benchmarks such as AudioBench~\citep{wang2025audiobenchuniversalbenchmarkaudio}, MMAU~\citep{sakshi2024mmaumassivemultitaskaudio}, and OmniBench~\citep{li_omnibench_2024} primarily test foundational tasks, with only the latter supporting full omni-modality. In contrast, \benchmark is grounded in chat-style interaction, reflecting how OVAs are deployed in real-world use. This setting demands deeper contextual understanding and flexible reasoning.

\begin{table*}[ht]
\centering
\resizebox{1.0\linewidth}{!}{%
\begin{tabular}{@{}l|ccccccccccc@{}}
\toprule
\multicolumn{1}{c}{\multirow{2}{*}{\textbf{Name}}} & \multirow{2}{*}{\textbf{Size}} & \multicolumn{3}{c}{\textbf{Acoustic Scene}} & \multicolumn{3}{c}{\textbf{Paralanguage}}        & \multicolumn{3}{c}{\textbf{Speaker Profile}}     & \multirow{2}{*}{\textbf{Avg. CA}} \\ \cmidrule(lr){3-11}
\multicolumn{1}{c}{}                      & \multicolumn{1}{c}{}                      & \textbf{VG}   & \textbf{SG}  & \textbf{CA}  & \textbf{VG} & \textbf{SG} & \textbf{CA} & \textbf{VG} & \textbf{SG} & \textbf{CA} &                         \\ \midrule
Human                                     & -                                         & 95.30          & 82.50         & 4.37         & 96.00        & 92.5        & 4.33        & 96.50        & 95.10        & 4.36        & 4.35                    \\
\midrule
\rowcolor[HTML]{D9E1F2}
\multicolumn{12}{c}{\textit{Open Source Models}} \\
\midrule
Mini Omni2                                & 7.0B                                      & 79.24         & 16.14        & 1.53         & 79.20        & 23.12        & 1.79        & 84.50        & 09.00        & 2.01        & 1.74                    \\
VITA 1.5                                  & 1.6B                                      & 78.12          & 16.50        & 2.60         & 81.14        & 34.57        & 2.56        & 88.60        & 14.20          & 3.01        & 2.69                    \\
VideoLlama2                                  & 7.0B                                      & 68.12          & 28.75        & 1.52         & 73.42        & 2.71        & 1.59        & 79.20        & 14.79          & 1.38        & 1.50                    \\
Baichuan-Omni                                  & 7.0B                                      & 76.15          & 34.37        & 1.90         & 77.24        & 32.71        & 2.25        & 84.20        & 16.20          & 2.01        & 2.02                    \\
Mini CPM                                  & 8.0B                                      & 87.62          & 35.35        & 2.87         & 89.14        & 39.28        & 2.35        & 88.40        & 25.40        & 2.90        & 2.69                    \\
Intern Omni                               & 8.7B                                      & 80.75          & 20.25        & 2.54         & 80.71       & 14.57        & 1.94        & 82.60       & 06.60        & 2.64        & 2.35                    \\
phi4 multimodal                           & 5.6B                                      & 81.24         & 23.12        & 2.26         & 79.14        & 33.57        & 2.63        & 84.57        & 12.40        & 2.48        & 2.44                    \\
Qwen 2.5 Omni                             & 7.0B                                      & 84.87         & 15.37        & 3.19         & 89.42        & 38.71        & 2.98        & 91.40       & 11.20        & 3.06        & 3.08   
\\
Qwen 2.5 Omni COT                             & 7.0B                                      & 83.50         & 24.50        & 3.27         & 88.28        & 26.71        & 3.00        & 88.80       & 18.00        & 3.33        & 3.19   
\\
\midrule
\rowcolor[HTML]{D9E1F2}
\multicolumn{12}{c}{\textit{Proprietary Models}} \\
\midrule
Gemini 2.5 Flash                               & -                                        & 89.50          & \textbf{59.00}         & 3.55         & 91.14       & 75.42       & 3.19        & \textbf{92.20}       & 65.60       & 3.64        & 3.44                    \\
Gemini 2.5 Pro                               & -                                        & \textbf{91.25}          & 54.75         & \textbf{3.65}         & \textbf{92.00}       & \textbf{77.42}       & \textbf{3.32}        & 91.60       & \textbf{71.60}       & \textbf{3.74}        & \textbf{3.56}                    \\\bottomrule
\end{tabular}
}
\caption{\small Performance breakdown of human and model responses on \benchmark across key skill domains. Visual Grounding (VG) and Speech Grounding (SG) evaluates the ability to perceive specific information in the modality needed for answering the question. Contextual Appropriateness (CA) evaluates the ability to produce contextually appropriate and accurate answers given the multimodal cues}
\label{tab:main_result}
\end{table*}

\noindent \textbf{Multi-Modal Inputs} Benchmarks like SD-Eval~\citep{ao2025sdevalbenchmarkdatasetspoken}, VoxDialog~\citep{cheng2025voxdialogue}, and S2S-Arena focus on speech-only inputs, targeting paralanguage or acoustic scene understanding in isolation. Others like Lyra SVQA~\citep{zhong2024lyraefficientspeechcentricframework} incorporate visual input but neglect paralinguistic cues. \benchmark is unique in requiring models to jointly interpret speech, vision, and background context, aligning with the OVA use case.

\noindent\textbf{Human Speech} Most existing chat benchmarks rely on synthetic speech, which current TTS systems struggle to render with accurate emotion or prosody~\citep{Wu2024LaughNC, Tang2023EmoMixEM}. In \benchmark, all spoken queries are recorded by professional voice actors to preserve natural paralinguistic signals. To validate this, we conduct a user study across 100 paralanguage-focused samples, comparing human-recorded speech to CosyVoice~\citep{du2024cosyvoice2scalablestreaming} and ElevenLabs TTS~\citep{elevenlabs2025}. Ten annotators rated speech on (1) attribute match and (2) naturalness. Human speech scored 4.6/4.5, compared to 2.4/2.1 (CosyVoice) and 3.1/3.3 (ElevenLabs), supporting the decision to use professional recordings.

\noindent\textbf{Confounders} Many existing benchmarks are susceptible to shortcut exploitation via textual or visual priors~\citep{kiela2021hatefulmemeschallengedetecting, goyal2017makingvvqamatter}. \benchmark introduces confounder pairs, where paralinguistic speech properties are inverted while keeping textual and visual input constant. This isolates the model’s ability to process non-verbal speech information and provides a more diagnostic and fine-grained evaluation framework. 

\section{Experiments}

\noindent\textbf{Evaluated Models} We evaluate a wide range of OLMs. The proprietary model assessed is Gemini-2.0-flash~\citep{google2024gemini}.For open-source OLMs we use Mini-Omni2~\citep{xie_mini-omni2_2024}, VideoLLama~\citep{zhang2023videollamainstructiontunedaudiovisuallanguage}, MiniCPM-o2.6~\citep{hu2024minicpmunveilingpotentialsmall}, Phi4-MM~\citep{microsoft2025phi4minitechnicalreportcompact}, VITA1.5~\citep{fu2025vita15gpt4olevelrealtime}, Baichuan Omni~\citep{li2025baichuanomni15technicalreport}, Intern Omni~\citep{chen2024internvlscalingvisionfoundation}.

\section{Main Results}

Table~\ref{tab:main_result} summarizes the performance of several proprietary and open-source OVAs on \benchmark. We highlight three key findings:

\begin{itemize}
    \item \textbf{\benchmark is challenging.} Although the tasks are straightforward for humans (average CA score: 4.33), the best-performing model (Gemini) only achieves 3.56—indicating that current OLMs struggle to integrate multi-modal cues, particularly non-verbal speech signals, even in seemingly simple scenarios.
    \item \textbf{Gap between proprietary and open-source models.} Gemini 2.5 Flash and Pro models outperform open-source models in their ability to ground in multi-modal inputs and provide  contextual responses. 
    \item \textbf{Speech grounding remains the bottleneck.} All models show relatively strong visual grounding, but consistently struggle to interpret non-verbal speech cues such as tone, emotion, and background sounds.
\end{itemize}

To better understand the sources of these limitations, we conduct a detailed analysis of Gemini 2.5 Pro, the best-performing model on \benchmark. We examine its behavior across core skill domains, focusing on how well it grounds responses in visual and speech cues, and identifying the types of errors that arise.

\begin{figure*}[t]
    \centering
    \begin{minipage}[t]{0.48\textwidth}
        \centering
        \includegraphics[width=\linewidth]{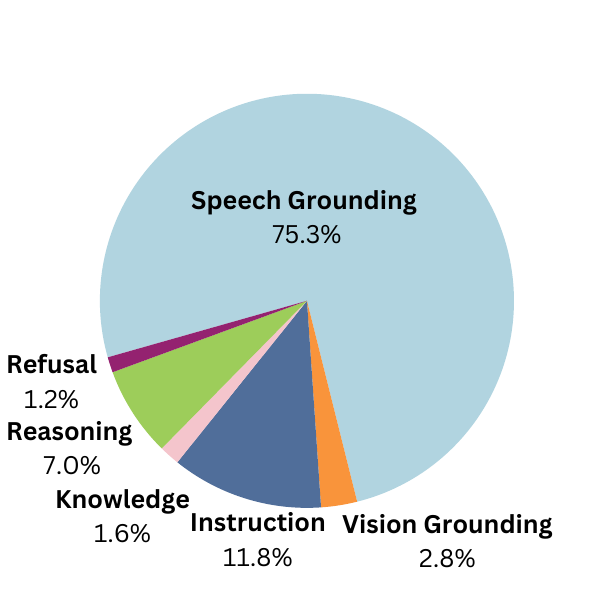}
        \caption{\small Distribution of error types in Gemini's responses on the MULTiVOX benchmark. The majority of errors (75.3\%) stem from speech grounding, indicating difficulty in interpreting non-textual auditory cues.}
        \label{fig:error_analysis}
    \end{minipage}%
    \hfill
    \begin{minipage}[t]{0.48\textwidth}
        \centering
        \includegraphics[width=\linewidth]{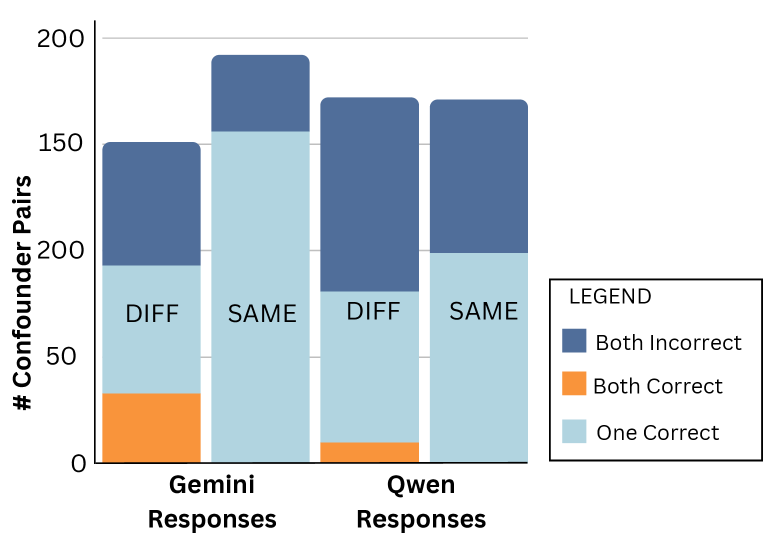}
        \caption{\small Distribution of model responses across confounder pairs. Bars indicate whether model answers were the same or different when speech cues were flipped. Ideally, answers should differ (left bar), but models often give identical responses (right bar), showing insensitivity to non-verbal speech cues.}
        \label{fig:confounder_analysis}
    \end{minipage}
\end{figure*}

\subsection{Where do models fall short?}

We analyze Gemini 2.5 Pro and Qwen 2.5 Omni performance across three skill domains: visual grounding, speech grounding, and multimodal reasoning. This breakdown highlights where current OLMs are reliable—and where they still fall short.

\subsubsection{Visual Grounding: A Strength}

Table~\ref{tab:main_result} shows the vision grounding scores for Gemini and Qwen 2.5 Omni. We find that these models demonstrate consistently strong visual grounding capabilities across tasks like object detection, scene understanding, and scene text recognition. These results indicate robust understanding of both fine-grained visual elements and broader scene context. While Qwen performs competitively with Gemini, Gemini performs better in acoustic scene category, where there are several tasks that require scene text understanding. Apart from this, most errors arise from under-specificity in open-ended scenarios - for instance, recognizing a “person holding game controllers” without identifying attributes like “retro game controllers”. This level of ambiguity is expected in real-world scenes and the high performance even in such conditions represents strong visual grounding capabilties.

\subsubsection{Speech Grounding: A Bottleneck}

In contrast to visual grounding strengths, our analysis reveals weaknesses in speech grounding capabilities (Table~\ref{tab:main_result}). We analyze the final responses generated by the model to evaluate if it is able to integrate audio characteristics in its responses. 

\noindent\textbf{Models hear the voice, but doesn’t ``recognizes'' the speaker profile} We check the model's ability to infer demographic characteristics (age and gender) from speech cues. Analysis of Qwen's responses show that the model relies on textual rather than acoustic content to identify speaker attributes in 30\% of cases. Additionally, the model rarely commits to definitive answers (16.0\% of cases), instead offering ambiguous answers (14.6\%), expressing uncertainty (14.4\%), or refusing to respond (65.6\%). In contrast to open-source models, we observe that Gemini consistently performs much better at perceiving speaker attributes such as age and gender. 

We further analyze the final responses to evaluate how well OLMs utilize this speech information in their final response. In case of Qwen, for questions requiring gender inference, the model provides neutral responses in 60\% of cases, with the remaining responses showing no significant bias toward either male or female speakers. Notably, when the model does make a gender inference, it appears to be influenced primarily by visual context and the question's content rather than speech characteristics. Gemini surprisingly shows similar trends, indicating that OLMs prefer a neutral response even if it can accurately infer their gender. Age-related inferences on the other hand reveal a stronger bias pattern, with both OLMs overwhelmingly favoring young-to-middle-aged adult in their responses regardless of the speaker's actual age. While Gemini is able to make accurate inferences regarding age, when generating final responses, this aspect is not integrated in its final responses. These errors have functional implications. For instance, these OLMs suggested potentially unsafe activities to speakers with children's voices in 20\% of the cases.

\noindent\textbf{Models struggle to utilize background sounds} In this category we test OLMs understanding of background music, sound and ambient noise. The evaluation of acoustic scene understanding reveals significant limitations, with Qwen achieving 15\% grounding score. Among these correct inferences, there is no significant different in performance among music recogntion and environmental sound recognition. Gemini performs much better at understanding background sound and music, achieving 54\% accuracy and we observe that it is also notably better at music understanding tasks. In questions with ambient noise, we observe that both OLMs demonstrate resilience to noise when processing queries. However, they still are limited in their capability to perceive noise. For instance, Qwen appears to rely predominantly on visual cues rather than acoustic features, as evidenced in approximately 39.5\% of cases where noisy environments were identified primarily through visual context (such as crowded airports).

\noindent\textbf{Models rely on textual cues for emotion understanding} While OLMs perform comparatively better in emotion understanding, we observe that they tend to rely on textual content rather than acoustic features. For instance Qwen explicitly relies on text in 71\% of cases and 26\% of the cases with Gemini. While emotions typically have a strong correlation with text, we deliberately introduced adversarial samples (Fig~\ref{fig:main_diagram} where textual and acoustic emotional cues conflict. Moreover, the confounder pairs have different emotions and we observe that Qwen only gets both emotions right in 27\% cases and Gemini in 50\% of the cases.  We observe similar performance trends in other paralanguage categories where there is an over-reliance on textual cues and the overall visual context.

\noindent \textbf{Limitation in spoken instruction following} We observe that open-source OLMs are limited in their spoken instruction following~\cite{chen2024omnixrevaluatingomnimodalitylanguage}, especially for instructions used in speech grounding where a grounding questions precedes the actual sample in our benchmark. Moreover, detailed explanations for answer could help in further detecting and penalizing modality shortcuts. For future work, we could consider using text modality as input to speech grounding questions for deeper analysis with instructions to explain its response.

\subsection{What causes these errors?}

We conducted a manual error analysis of Gemini's outputs on \benchmark to identify underlying failure patterns. Fig.~\ref{fig:error_analysis} shows that perception errors dominate, accounting for 75.3\% of all failures, primarily reflecting the model's inability to ground to speech cues. Reasoning failures constitute 7.0\% of errors, indicating that even when the model successfully perceives multimodal inputs, it struggles to effectively integrate this information to generate appropriate responses. Instruction understanding failures represent a similarly significant error category, where the model defaults to describing visual content rather than addressing the intended query. \emph{The overwhelming majority of speech perception errors indicates a clear bottleneck: improving speech perception capabilities is essential for building effective OVAs in multimodal contexts.}

\subsection{Do models really listen to speech cues?}

To assess whether models are truly leveraging speech cues, we analyze their responses across confounder pairs, focusing on the top two performers: Gemini 2.5 Pro and Qwen2.5-Omni (Fig~\ref{fig:confounder_analysis}). Without considering confounders, both models appear moderately accurate, getting around 50\% of responses correct across these pairs. However, this aggregate accuracy can be misleading. When we examine whether models actually change their answers in response to flipped speech cues, we find that in a majority of the confounder pairs (57\% for Gemini, 51\% for Qwen), the model outputs are paraphrases, i.e., they show NO FLIP in answer. Crucially, within those NO FLIP cases, most instances with one correct answer suggest that the correctness arises from chance or visual/textual bias, not from grounding in speech. This indicates that, despite non-trivial accuracy, models are largely ignoring non-verbal speech cues when answering.


\section{Conclusion}

We introduce \benchmark, the first benchmark designed to evaluate Omni Language Models (OLMs) as omni-modal voice assistants (OVAs) that integrate speech and vision for context-aware reasoning. Unlike prior benchmarks that rely on synthetic speech or focus only on unimodal cues, \benchmark includes 1000 professionally recorded, human-spoken questions paired with images or videos, emphasizing paralinguistic signals like tone, emotion, and background noise. A key innovation is the use of confounder pairs—speech variants with identical text and visuals—to ensure models attend to speech beyond surface cues. \benchmark enables fine-grained diagnosis across speech, vision, and general reasoning skills. Evaluation of 10 state-of-the-art models shows that, while visual grounding is robust, speech grounding remains a significant bottleneck. Our benchmark will be open-sourced to support the development of truly multimodal voice assistants.

\section*{Limitations}

\begin{itemize}
    \item We limit ourself to questions to the english language, extending to multilingual settings is an important future direction to assess OLM generalization across languages.
    \item In this work, we limit our evaluation to the content of the speech outputs and not the speech quality of the output, such as naturalness and appropriateness. Evaluating speech synthesis and conversational prosody is an important but orthogonal direction left for future benchmarks.
\end{itemize}

\bibliography{custom}

\appendix

\section{Appendix}
\label{sec:additional}

In the Appendix, we provide:
\begin{enumerate}
    \item Section~\ref{sec:other_dataset_details}: Other Dataset Details
    \item Section~\ref{sec:annotator_details}: Annotator Details
    \item Section~\ref{sec:judge_details}: LLM-as-a-judge Details
    \item Section~\ref{sec:mos_test}: Voice Quality Assessment
\end{enumerate}

\section{Other Dataset Details}
\label{sec:other_dataset_details}

Here we detail the categories and sub-categories present in our benchmark. 

\subsection{Acoustic Scene Understanding}

\textbf{Background Music Understanding:} These tasks require the model to detect and interpret the presence and nature of background music in spoken queries. This includes genre classification, mood inference from music, and distinguishing music from speech or noise.

\textbf{Sound Event Recognition:} This task evaluates the model’s ability to detect and categorize discrete, identifiable audio events (e.g., dog barking, glass breaking, door closing) that occur within the auditory scene alongside spoken content.

\textbf{Ambient Environment Sound:} Models are tested on their capacity to recognize broader acoustic environments (e.g., airport, cafe, subway) based on background audio cues. We construct these scenes using the MS-Noise~\citep{nachmani2024spokenquestionansweringspeech} dataset, overlaying clean speech with environmental recordings at a signal-to-noise ratio (SNR) of -2 dB to simulate challenging real-world conditions.

\subsection{Paralanguage Understanding}

\begin{figure*}[t]
    \centering
    \begin{minipage}[t]{0.48\textwidth}
        \centering
        \includegraphics[width=\linewidth]{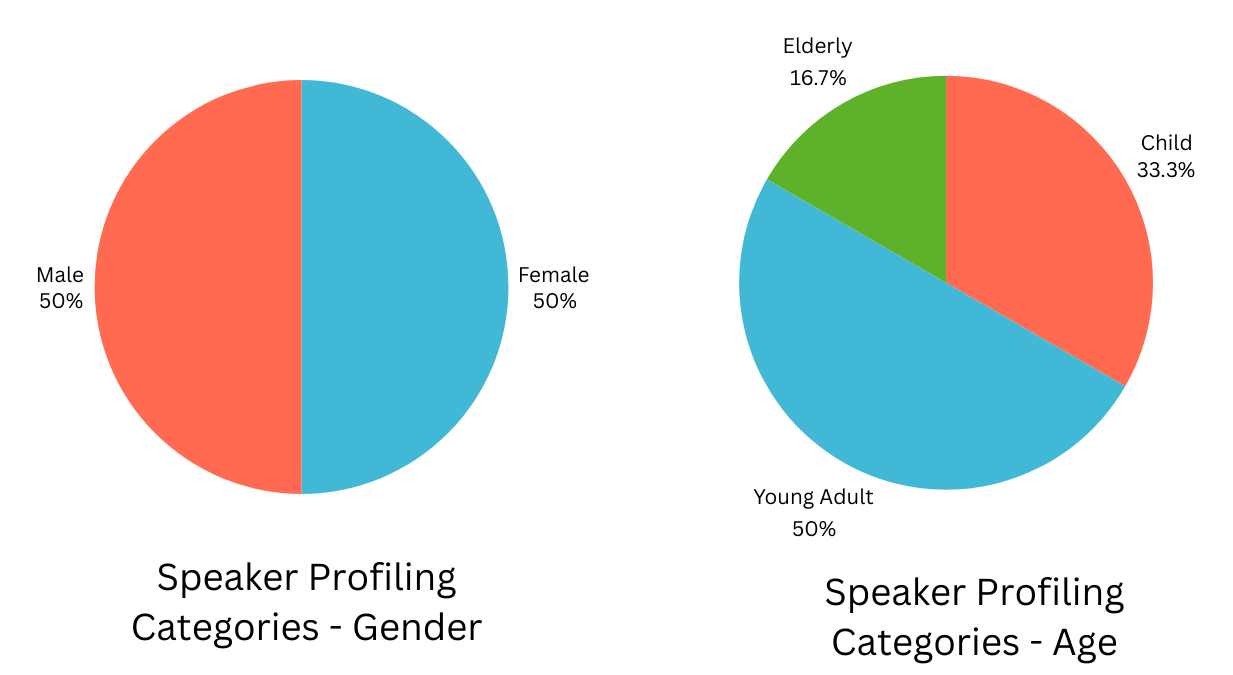}
        \caption{Pie charts showing different Speaker Profile categories in terms of Gender and Age}
        \label{fig:speaker_profiling}
    \end{minipage}%
    \hfill
    \begin{minipage}[t]{0.48\textwidth}
        \centering
        \includegraphics[width=\linewidth]{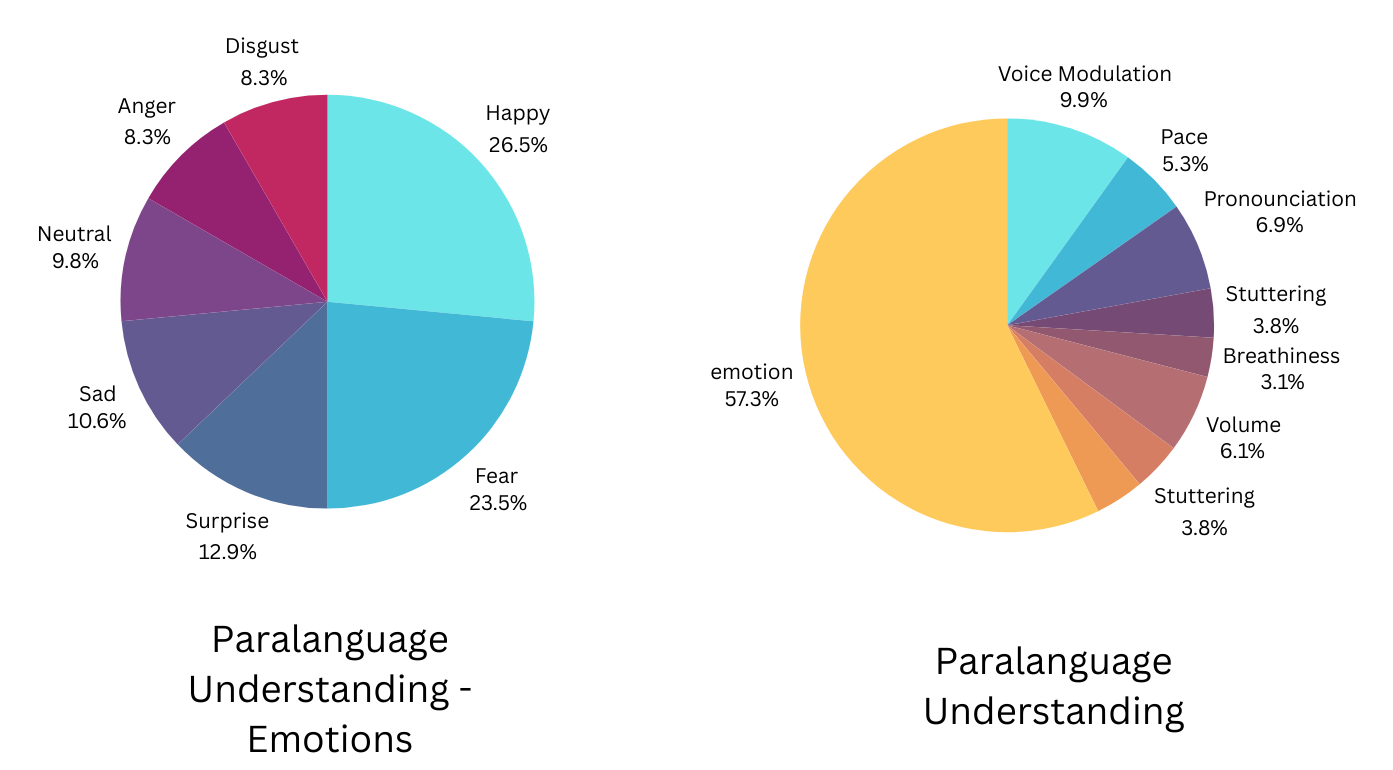}
        \caption{Pie charts showing different Paralanguage Understanding Categories and different Emotion Categories in the benchmark}
        \label{fig:paralanguage_understanding_dist}
    \end{minipage}
\end{figure*}

Detailed distribution of the categories here are listed in Fig~\ref{fig:paralanguage_understanding_dist}

\textbf{Emotion:} The task focuses on identifying the emotional state of the speaker as conveyed through prosodic features (e.g., pitch, energy), independent of lexical content.

\textbf{Voice Modulation:} Evaluates the model’s sensitivity to dynamic vocal variations such as emphasis, intonation, and expressiveness, which can affect intent or meaning.

\textbf{Pronunciation:} Tasks involve recognizing deviations from standard pronunciation, which may signal emotion, emphasis, or speaker background.

\textbf{Volume:} Assesses the ability to perceive and reason about loudness cues, which can convey urgency, emotion, or social context.

\textbf{Pace:} Evaluates how well the model understands speech rate—e.g., rushed versus slow delivery—as a cue to emotional or cognitive states.

\textbf{Stuttering:} Measures recognition and interpretation of speech disfluencies, including repeated sounds or syllables, as part of speaker modeling or intent understanding.

\textbf{Breathiness:} Focuses on detecting breathy voice quality, which may indicate fatigue, emotion, or affective state.

\subsection{Speaker Profiling}

\textbf{Biological Gender Estimation:} The task is to estimate the speaker’s biological gender based solely on voice characteristics, controlling for content and visual input.

\textbf{Age Group Estimation:} Models must infer the age group (e.g., child, adult, elderly) of the speaker using acoustic cues.

\section{Annotator Details}
\label{sec:annotator_details}

\subsection*{1. Annotator Composition}
We formed a panel of six domain experts for our dataset creation and filtering process and our dataset review process. The panel consisted of four Ph.D students pursuing speech and audio-visual research and two MS students having research in speech and audio processing. The expertise of all domain experts is evidenced by their research publications and contributions to the field. 

\subsection*{2. Meetings to decide question creation process}
All six annotators had three 2-hour online meetings to discuss the question and corresponding answers creation process to reach a consensus of the pipeline to be followed for dataset creation. The online meetings covered these details:

\begin{itemize}
    \item \textbf{Multimodal Question-Answering Foundations:} Aligning nonverbal audio cues (e.g., tone, background sounds) with visual context (e.g., scene imagery).

    \item \textbf{Confounder Pair Design:} Generating minimal audio variants that invert answers (e.g., adding subtle noise to flip ``quiet'' vs. ``noisy'').

    \item \textbf{Annotation Platform \& Guidelines:} Hands-on use of our custom interface and details about input/output formats, edge cases, and scoring rubrics.
\end{itemize}

\noindent Following the online meetings, annotators jointly labeled 50 pilot samples and attained $\geq 90\%$ inter-annotator agreement before proceeding to full‐scale work.

\subsection*{3. Question Creation Process}
Each annotator followed a three-step pipeline:

\begin{itemize}
    \item \textbf{Scenario Drafting:} Write a conversational prompt targeting modality cues (e.g., ``Can you tell if this room is too loud for a conference call?'').
    \item \textbf{Confounder Generation:} Create a paired version of the prompt with the same text and visuals but alter one audio attribute (e.g., add background machinery noise).
    \item \textbf{Media Pairing:} Select or synthesize matching audio and visual assets from our curated libraries to illustrate both the original and confounded conditions.
\end{itemize}

\noindent This ensured every question/confounder pair isolated the intended cue and prevented shortcut learning by downstream models.

\subsection*{4. Answer Creation Process}
\label{sec:answer_creation}
For each question instance, annotators produced:

\begin{itemize}
    \item \textbf{Reference Answer:} A concise response directly addressing the prompt (e.g., ``No, it’s quiet enough for clear conversation'').
    \item \textbf{Rationale Statement:} A brief explanation linking the critical cue to the answer (e.g., ``The low ambient noise level confirms a silent office setting'').
\end{itemize}

\noindent These reference answers and rationales formed the ground truth for our GPT-4--based judgment of model responses.

\subsection*{5. Annotation Criteria \& Quality Control}
All questions and answers were crafted according to these overarching principles:

\begin{itemize}
    \item \textbf{Clarity \& Brevity}: Simple, conversational language devoid of syntactic complexity.
    \item \textbf{Modality Isolation}: Exactly one audio or visual ``hook'' per item, ensuring focused evaluation.
    \item \textbf{Balanced Distribution of Skills}: Even distribution of questions across different skills.
\end{itemize}

\section{LLM-as-a-Judge Details}
\label{sec:judge_details}

\begin{enumerate}
\item \textbf{Human vs. LLM Evaluation Experiment}: To validate whether a large language model (LLM) could reliably substitute for our expert reviewers, we conducted a blind evaluation on a sample of 300 question-answer pairs drawn evenly from our benchmark. Each pair was independently graded on a 1-5 scale by:

\begin{itemize}
    \item Three domain experts from our panel following a predecided grading criteria
    \item An \textit{LLM-as-a-Judge}, implemented via a single GPT-4.1-mini call per sample.
\end{itemize}

\noindent We computed average scores for each item under both conditions and measured inter-rater agreement using Cohen's $\kappa$. Across all 300 items, human--human agreement averaged $\kappa = 0.82$, while human--LLM agreement reached $\kappa = 0.78$, indicating the LLM’s judgments were almost as consistent with the experts’ as the experts were with one another. Overall the LLM’s ratings fell within one point of the experts’ in 92\% of cases.

Given these results, we adopted the \textit{LLM-as-a-Judge} for large-scale scoring in subsequent experiments, leveraging its efficiency without materially sacrificing quality.
\item \textbf{LLM-as-a-Judge Criteria}:

We present each question to the LLM-as-a-Judge by supplying both the corresponding speech and visual inputs. The model is then asked to rate the provided answer on a five-point scale according to the following criteria:

\begin{itemize}
  \item \textbf{Score 1:} The response is often off-topic or incorrect and fails to recognize or use the specified speaker characteristic.
  \item \textbf{Score 2:} The response occasionally addresses the prompt but handles the speaker characteristic inconsistently or superficially.
  \item \textbf{Score 3:} The response shows a basic understanding of intent, with partial integration of the speaker characteristic but lacks depth or precision.
  \item \textbf{Score 4:} The response delivers relevant and mostly accurate content that usually incorporates the speaker characteristic, with only minor lapses.
  \item \textbf{Score 5:} The response consistently produces accurate, context-rich answers that fully and effectively integrate the speaker characteristic.
\end{itemize}

\end{enumerate}

\section{Voice Quality Assessment}
\label{sec:mos_test}

We generated synthetic audio for 100 benchmark questions using two text-to-speech systems: CosyVoice (open source)~\citep{du2024cosyvoice2scalablestreaming} and ElevenLabs~\citep{elevenlabs2025} (commercial). We also recorded these questions by professional voice actors.

\noindent To compare these renderings, we conducted a Mean Opinion Score (MOS) Test on a random subset of 100 \emph{paralanguage-focused} queries (as in Sec.~3.4). Ten expert annotators rated each audio on two dimensions, using a 5-point scale:  
\begin{itemize}
  \item \emph{Attribute match}: How accurately the intended paralinguistic cue (e.g., emotion, background noise) was conveyed.  
  \item \emph{Naturalness}: The overall human-likeness of the voice.  
\end{itemize}

\noindent The MOS results were as follows: Professional human recordings: 4.6 (attribute match) / 4.5 (naturalness), CosyVoice TTS: 2.4 / 2.1, ElevenLabs TTS: 3.1 / 3.3  .

\noindent These findings confirm that, while modern TTS can approximate certain prosodic features, professional voice actors remain far superior in both fidelity and naturalness, which led us to use human-recorded queries by preofessional voice actors throughout the benchmark.

\section{Voice Data Collection}
\label{sec:voice_data_collection}
To ensure natural and expressive speech, we employed four professional voice actors, four male and two female, contracted via Fiverr. Each actor was compensated according to the standard rates listed on the platform. This approach allowed us to capture high-quality, emotionally varied recordings with realistic prosody and delivery across all tasks. Our institution’s Institutional Review Board (IRB) has granted approval for this data collection.

\section{Additional Details: Auxiliary}

\noindent\textbf{Compute Infrastructure:} All our experiments are conducted on a single NVIDIA A6000 GPU. No training is required, and depending on the downstream task, a single inference run on a benchmark requires anywhere between 30 minutes to 2 hours. 

\noindent\textbf{Implementation Software and Packages:} We use the official implementation of the OLMs we benchmark. For LLM-as-judge implementation, we utilize the official OpenAI APIs.

\noindent\textbf{Potential Risks:} We manually curate all our questions to avoid any potential harmful or biased samples.

\end{document}